\documentclass[notitlepage,showpacs,prl,reprint]{revtex4-1}

\usepackage{amsmath}
\usepackage{epsf}
\usepackage{graphicx}

\usepackage{dcolumn}
\usepackage{bm}

\usepackage{amssymb}
\usepackage{array}
\usepackage{varioref}
\usepackage{wrapfig}
\usepackage{etoolbox}
\usepackage{color}

\providecommand{\nn}{\nonumber}

\providecommand{\bv}[1]{\bm{\mathrm{#1}}}

\providecommand{\W}{\Omega}
\providecommand{\q}{\bv{q}}

\renewcommand{\k}{\bv{k}}

\providecommand{\ef}{\varepsilon_F}
\providecommand{\vf}{v_F}

\providecommand{\kf}{k_F}

\providecommand{\kf}{k_F}

\renewcommand{\q}{\bv{q}}

\providecommand{\gb}{\bar{g}}

\providecommand{\tp}{2\pi}

\usepackage{feynmp-auto}
\usepackage[caption=false]{subfig}
\usepackage{etoolbox}
\usepackage{mdframed}
\usepackage{ulem}

\usepackage{graphicx}

\begin{document}
\title{Dynamical susceptibility of a near-critical non-conserved order parameter and $B_{2g}$ Raman response in Fe-based superconductors}

\author{Avraham Klein}
\affiliation{School of Physics and Astronomy, University of Minnesota, Minneapolis. MN 55455}
\author{Samuel Lederer}
\affiliation{Department of Physics, Massachusetts Institute of Technology, Cambridge MA 02139}
\author{Debanjan Chowdhury}
\affiliation{Department of Physics, Massachusetts Institute of Technology, Cambridge MA 02139}
\author{Erez Berg}
\affiliation{Department of Physics, University of Chicago, Chicago IL 60637}
\author{Andrey Chubukov}
\affiliation{School of Physics and Astronomy, University of Minnesota, Minneapolis. MN 55455}

\begin{abstract}
  We analyze the dynamical response of a two-dimensional system of itinerant fermions coupled to a scalar boson $\phi$, which undergoes a continuous transition towards nematic order with $d-$wave form-factor.  We consider two cases: (a) when $\phi$ is a soft collective mode of fermions near a Pomeranchuk instability, and (b) when it is an independent critical degree of freedom, such as a composite spin order parameter near an Ising-nematic transition. In both cases, the order-parameter is not a conserved quantity and the $d-$wave fermionic polarization $\Pi (q, \W)$ remains finite even at $q=0$.  The polarization $\Pi (0, \W)$
  has similar behavior in the two cases, but the relations between  $\Pi (0, \W)$ and the bosonic susceptibility $\chi (0, \Omega)$ are different, leading to different forms of $\chi^{\prime \prime} (0, \Omega)$, as measured by Raman scattering.
  We compare our results with  polarization-resolved Raman data for the Fe-based superconductors FeSe$_{1-x}$S$_x$, NaFe$_{1-x}$Co$_x$As  and BaFe$_2$As$_2$.
  We argue that the data for FeSe$_{1-x}$S$_x$ are
  well
  described within Pomeranchuk scenario, while
   the data for NaFe$_{1-x}$Co$_x$As  and BaFe$_2$As$_2$
  are better  described within
  the ``independent'' scenario involving a composite spin order.
\end{abstract}

\maketitle
{\it Introduction.-}~ The behavior of strongly correlated fermions in the vicinity of a quantum critical point (QCP) is one of the most fascinating problems in many-body physics.  A traditional way to treat the physics near a QCP is to study an effective low-energy  model in which itinerant fermions are coupled to near-critical fluctuations of a bosonic order parameter \cite{Abanov2003}. The boson can be a collective mode of electrons, as in the case of a Pomeranchuk instability, or an independent degree of freedom (e.g., a phonon). In both cases, the  boson-fermion coupling affects the bosonic dynamics. This effect is encoded in the fermionic polarization $\Pi (\q, \W)$, which in turn is related to the bosonic susceptibility $\chi (\q,\W)$. Previous studies of $\chi (\q,\W)$  \cite{Hertz1976,Millis1992,Millis1993,Altshuler1994,Abanov2003,Metlitski2010b,Maslov2010,Sachdev2012,Lee2017} focused primarily on the range $\W \ll v_F q$ ($v_F$ is the Fermi velocity), in which the scaling behavior holds in critical theories with a dynamical exponent $z >1$  .
However, several experimental probes, most notably polarization-resolved Raman scattering, analyze $\chi (\q, \W)$ in the opposite limit of vanishing $q$ and finite $\W$ ~\cite{Devereaux2007}.  The same regime has been probed in  Quantum-Monte-Carlo studies~\cite{Schattner2016}.
 If the order parameter is conjugate to a conserved quantity, e.g. the total fermion number-density or the total spin, the fermionic polarization $\Pi (\q, \W)$ vanishes identically by the conservation law at $q=0$ and, by continuity, is
small at $\W \gg v_F q$. However, if the order parameter is conjugate to a quantity that is not constrained by conservation laws,  $\Pi (0, \W)$ does not have to vanish and may
 give rise to a nontrivial frequency dependence of $\chi (0,\W)$.

In this letter we report the results of our study of $\Pi (0, \W)$ and $\chi (0,\W)$ for a system of fermions in two spatial dimensions, coupled to fluctuations of a (charge) nematic order parameter, $\phi$, with a $d-$wave form-factor. If
 $\phi$  is a collective mode of fermions, the model describes an itinerant fermionic system near a Pomeranchuk instability.
  If $\phi$ is a separate degree of freedom, it softens on its own, but fermions still affect the critical behavior.
 We will see that $\Pi (\q=0, \W)$ is the same in both cases, but the relationship between $\Pi (0, \W)$ and the uniform dynamic bosonic susceptibility $\chi (0,\W)$ are different, leading to different predictions for Raman experiments. In the case of a Pomeranchuk instability,
 $\chi (q,\W)$ is necessarily proportional to $\Pi (q,\W)$ for all momenta and frequencies,  and we argue that $\chi (0,\W) \approx -\Pi (0,\W)$. On the other hand, when $\phi$ is an independent near-critical order parameter, the bosonic susceptibility nearly coincides with the susceptibility of the $\phi$ field, and $\Pi (0,\W)$  enters $\chi (0,\W)$  as a bosonic self-energy:
  $\chi^{-1} (0,\W) = \chi_0^{-1} (0,\W)-
  {\bar g} \Pi (0,\W)$, where $\chi_0 (0,\W)$ is the dynamic susceptibility of the $\phi$ field in the absence of interactions with fermions,  and ${\bar g}$ is the fermion-boson coupling (defined below).

We argue that at the QCP, the imaginary part of the polarization operator $\Pi^{\prime \prime} (0, \W)$ scales as  $\W^{1/3}$ at low frequencies and as $\W$ at somewhat higher frequencies, which are still smaller than the Fermi energy, $\ef$.   Away from the QCP, when the nematic correlation length $\xi$ is large but finite, the $\W^{1/3}$ dependence is replaced by $\Pi^{\prime \prime} (0, \W) \propto \W \xi^2$ below $\Omega \propto \xi^{-3}$.
 The real part   $\Pi^\prime (0,\W)$ remains finite at $\W \to 0$, but $\Pi (0,0)$ evolves rather strongly with $\xi$ in systems with
 a small Fermi momentum $k_F$. Technical details of our analysis are presented in~\cite{Klein2017a}.

We compare our results with the Raman data for the
  Fe-chalcogenide system FeSe$_{1-x}$S$_x$ \cite{Massat2016,Blumberg2017} and the  Fe-pnictides BaFe$_2$As$_2$ \cite{Yang2014} and NaFe$_{1-x}$Co$_x$As \cite{Thorsmolle2016}, which all display nematic order in some range of temperature and doping. For these and related systems two different electronic scenarios for nematicity have recently been put forward~\cite{Fernandes2014}. One scenario is that nematicity is associated with a composite spin order~\cite{Fernandes2012}, while
 another is that nematic order is a Pomeranchuk order in the charge channel with a $d-$wave form factor. ~\cite{Gallais2016a,Gallais2016,Classen2017,Onari2012}

 For FeSe and FeSe$_{1-x}$S$_x$, we find good agreement with the Pomeranchuk scenario, consistent with the fact that magnetic order in these systems does not develop down to $T = 0$. For  NaFe$_{1-x}$Co$_x$As and BaFe$_2$As$_2$ we find an agreement with the composite spin (Ising-nematic) scenario of nematicity. This is consistent with the fact that in NaFe$_{1-x}$Co$_x$As nematic and stripe magnetic ordering temperatures differ by only a few Kelvin and follow each other as  functions of doping.

\begin{figure}
  \centering
  \subfloat[]{\includegraphics[width=0.95\hsize,clip,trim=0 175 0 175]{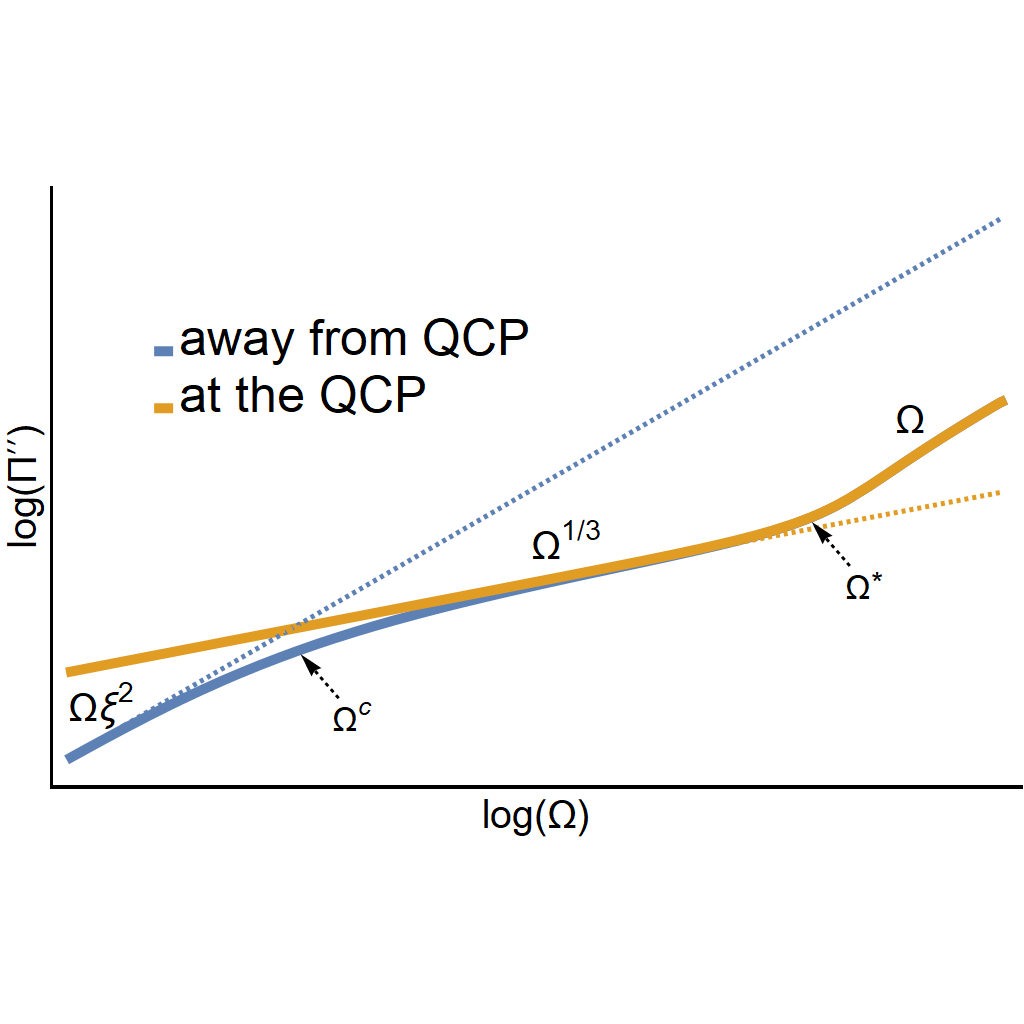}}
  \caption{
     The d-wave fermion polarization $\Pi^{\prime \prime} (q, \W)$ at $q=0$ for a model of itinerant fermions near an Ising nematic quantum critical point.
       We use log-log scale to highlight different power-law regimes.
      At very low frequencies, $\Pi^{\prime \prime} (0, \W)$ scales as $\W \xi^2$.  Above a crossover frequency $\Omega^c \propto \xi^{-3}$,  $\Pi^{\prime \prime} (q, \W)$ becomes universal (independent of $\xi$) and has different forms in two regimes, depending on the ratio $\W/\W^*$, where $\W^* \ll \ef$ is defined after Eq. (\ref{eq:pi-bare-wrong}).
      For $\W \ll \W^*$, $\Pi^{\prime \prime} (0,\W) \propto |\W|^{1/3}$. For $\W \gg \W^*$,
        $\Pi^{\prime \prime} (0,\W) \propto |\W|$.   At even higher frequencies (not shown) $\Pi^{\prime \prime} (0,\W)$ passes through a maximum and decreases.}
  \label{fig:susc}
\end{figure}

{\it The model.-}~
We consider a clean two-dimensional system of itinerant fermions with a circular Fermi surface (FS) specified by Fermi momentum $\kf$ and Fermi velocity $v_F$, coupled to a scalar boson $\phi (q)$, which undergoes a continuous transition towards charge nematic order with $d-$wave form factor. The field $\phi$ is coupled to the $d-$wave component of fermionic density as
\begin{equation}
  \label{eq:coupling-def}
  H_I = g \sum_{\k,\q}f(\k)\phi(\q)\psi^{\dagger}(\k+\q/2)\psi(\k-\q/2),
\end{equation}
where $g$ is a coupling constant and $f(\k)$ is a momentum dependent vertex with $d-$wave symmetry (e.g., $f(\k) = \cos {k_x} - \cos{k_y}$).
Near the FS  $f(\k) \approx f(\theta)$, where $\theta$ is an angle along the FS.  This model  has been discussed extensively in the regime where the characteristic bosonic frequency $\W$ is much smaller than $v_F q$, with $q$ the characteristic bosonic momentum~\cite{Altshuler1994,*Lee2017,*Metlitski2010b,*Sachdev2012,*Rech2006,*Maslov2010,*Lawler2007,*Chubukov2006,*She2017,Holder2015}. In standard treatments of this regime, the  boson propagator is
\begin{equation}
  D (\q,\W) = \frac{\chi_0}{\xi^{-2}_0 +{\q}^2  -\Omega^2/c^2 +\gb \Pi  (\q,\W)}.
  \label{eq:chi}
\end{equation}
Here $\gb= g^2 \chi_0$ is the effective coupling constant (assumed much smaller than $\varepsilon_F\equiv v_Fk_F/2$), $\xi_0$ and $c$ are the bosonic correlation length and velocity at $g=0$, i.e., in the absence of coupling to fermions, and $\Pi (q,\W)$ is the particle-hole polarization bubble, given by
\begin{equation}
  \label{eq:pi-1-loop}
  \gb\Pi(\q,\W) = -\frac{\gb k_F^2}{4\pi\ef} \left(\langle f^2 \rangle + i f^2({\hat \q}')\frac{|\W|}{\vf |\q|}\right),
\end{equation}
where
$\ef = v_f k_F/2$,
 $\langle f^2(\theta) \rangle = \int \frac{d\theta}{\tp}f^2(\theta)$, and ${\hat \q}' = {\hat z} \times {\hat \q}$, ${\hat z}$ is a unit vector in the direction perpendicular to the 2D plane~\cite{Oganesyan2001,Chubukov2003}.
 The constant term in $\gb\Pi(\q,\W)$ accounts for the difference between $\xi_0$ and the actual nematic correlation length $\xi$:
 $ \xi^{-2} = \xi_0^{-2} - \gb k_F^2\langle f^2 \rangle/(4\pi\ef) $.
The form of $D(\q,\W)$ at $\W \ll v_F |\q|$  determines the fermionic self-energy on the FS: at the QCP, $\Sigma (\W, \theta) \propto |f(\theta)|^{4/3} |\W|^{2/3} \omega^{1/3}_0$, where $\omega_0 \sim {\bar g}^2/\ef$.

Our goal is to obtain $\Pi (\q,\W)$ at $T=0$ in the opposite limit of $\q=0$ and finite $\W$. We obtain $\Pi (0,\W)$ first along Matsubara axis $\W=i\W_m$, and then convert to real $\W$.  For free fermions, $\Pi (0,\W_m)$ vanishes for arbitrary $f(\theta)$ because the
density of fermions at each momentum
is separately conserved.  At a finite $\gb$, this is generally not the case. We evaluate  $\Pi (0,\W_m)$ to leading order in $\gb$ by  computing the two-loop
 Maki-Thompson
diagrams for the d-wave particle-hole bubble, along with the Aslamazov-Larkin diagrams, which contribute at the same order (see Ref. \cite{Klein2017a} for details).
For a constant form-factor ($f(\theta)=1$), these diagrams
cancel exactly, and the cancellation can be traced to the Ward identity for fermion number conservation \cite{Baym1961}. For a non-conserved order parameter the
diagrams do  not cancel. Evaluating the diagrams, we obtain
\begin{flalign}
  \label{eq:pi-bare-wrong}
  \gb\Pi (\q = 0, \W_m) &=
  \left(\frac\gb{v_F}\right)^2
  \langle f_2\rangle \times \nonumber \\
  &\left[A \left(k_F \xi\right)
     -
 C \left(|\Omega_m|;k_F\xi \right) \right].
\end{flalign}
Here $\langle f_2 \rangle = \langle f^2f'^2+\frac{1}{2}f^3f''\rangle < 0$, and the functions $A$ and $C$ are
\begin{align}
  \label{eq:finite-A}
  A\left(k_F\xi\right) &= 1 - (\kf\xi)^{-1}\tan^{-1}(\kf\xi)  \\
  &\simeq 1 -(\pi/2)(k_F\xi)^{-1} \quad(k_F \xi \gg 1), \nn
\end{align}
and
\begin{equation}
  \label{eq:finite-C}
   C \left(|\Omega_m|;k_F\xi \right)   \sim \left\{
    \begin{array}{ll}
      \frac{\gb |\Omega_m|}{\ef^2}(k_F\xi)^2 & |\Omega_m| \ll \Omega^c \\
      \left(\frac{\gb |\Omega_m|}{\ef^2}\right)^{1/3} & \Omega^c \ll |\Omega_m| \ll \Omega^* \\
\frac{v_F}{c}      \frac{|\Omega_m|}{\ef} & \Omega^* \ll |\Omega_m|\ll\ef\\
    \end{array}\right..
\end{equation}
The characteristic frequencies in (\ref{eq:finite-C}) are
 $\W^c=\ef^2\gb^{-1} (k_F\xi)^{-3}$ and $\W^*=\sqrt{\gb\ef(c/v_F)^3 }$. The frequency
 $\Omega^c$ separates Fermi liquid and QC behavior, and $\W^*$ marks where Landau damping starts to dominate over the bare $\W^2_m/c^2$ term of Eq. \ref{eq:chi}.
 Eqs. (\ref{eq:finite-A}) and ({\ref{eq:finite-C}) are valid when $\Omega^c \ll \Omega^*$, i.e. $k_F\xi \gg \sqrt{(\ef /\gb)(v_F/ c)}$,
which is always satisfied close enough to the QCP. The various dynamical regimes in real frequency, both at and near the QCP are depicted qualitatively in Fig. \ref{fig:susc}.

The perturbative calculation leading to Eqs. (\ref{eq:pi-bare-wrong})-(\ref{eq:finite-C}) is only valid when
$|\Sigma (\W_m)| \ll |\W_m|$, which
holds for $|\W_m| \gg\omega_0$. At smaller $\W_m$ higher-order self-energy and vertex corrections
cannot be neglected {\it a priori}. We computed $ \gb\Pi (0, \W_m)$ at $\W_m \ll \omega_0$ by inserting a series of self-energy and vertex corrections and found that, to leading order in $\gb/\ef$, these higher order contributions cancel each other, such that Eq. (\ref{eq:pi-bare-wrong}) remains unchanged \footnote{The cancellation can be traced to the fact that the relevant vertex corrections to diagrams
 are the same as for the density-density polarization, and they cancel out with self-energy insertions by a Ward identity. This reasoning is very similar to the one that a fermionic residue cancels out between self-energy and vertex corrections in the expression for the optical conductivity \cite{Maslov2017}}.

{\it Nematic susceptibility}~~~
 When $\phi$ represents a collective mode of fermions near a d-wave Pomeranchuk instability, the
 nematic susceptibility is
   well approximated by
   $\chi_{Pom} (0,\W) = -\Pi (0,\W)/(1 + \gb\xi_0^2 \Pi (0,\W))$ where $\gb\xi_0^2$ is a four-fermion interaction in the d-wave channel
   \footnote{To satisfy the Stoner-like criterion for such a transition, we must take $\gb \xi_0^2 \sim \ef /k_F^2$, and therefore either relax our weak coupling assumption or assume a fermionic interaction of long but finite range $\xi_0$.}.
   For $\W \ll v_F q$, one can verify that  $\chi_{Pom}(\q,\W)$  has the same form as $D(\q, \W)$ in Eq. (\ref{eq:chi}).  For $\W \gg v_F q$, the denominator of $\chi_{Pom}$ is no longer singular, and its behavior is qualitatively that of $\Pi$, $\chi_{Pom} \sim -\Pi(0,\W)$.
   When $\phi$ can be considered  an independent, near-critical bosonic field, the full bosonic susceptibility $\chi_{ind} (0,\W)$ predominantly comes from the $\phi$ field, i.e.,
 $\chi_{ind} (0,\W) \approx D(0,\W)$, where $D$ is given by Eq. (\ref{eq:chi}).
 The behavior of $\chi_{ind} (0,\W)$ then  depends on the scale $\xi_0$ as well as $\xi$. This opens an avenue to distinguish the ``Pomeranchuk" and ``independent" scenarios, by comparing low-energy properties of the real and imaginary parts of their corresponding susceptibilities. We concentrate on two such properties: the dependence of $\chi'_{\xi}(0,\W\to 0)$ on $\xi$, specifically on the behavior of $\delta\chi' = \chi'_{\xi=\infty}(0,\W\to 0)-\chi'_{\xi}(0,\W\to 0)$, and the slope of the imaginary part of the susceptibility $\Gamma = \left.\chi^{\prime \prime}(0,\W)/\W\right|_{\W\to 0}$.

In the Pomeranchuk case, converting Eqs (\ref{eq:finite-A}) and (\ref{eq:finite-C}) to real frequencies we find
\begin{align}
  \label{eq:pom-scaling}
 \delta \chi^\prime_{Pom} \propto \xi^{-1}, \quad\Gamma_{Pom}\propto \xi^2.
\end{align}
In the independent case, to leading order in ${\bar g}/\ef$, $\chi^\prime_{ind} ( \W) = \chi_0 \xi^2_0$, and $\chi^{\prime \prime}_{ind} (\W) =-\chi_0 {\bar g} \xi^4_0 \Pi^{\prime \prime} (0,\W)$, where we recall that $ \xi^{-2} = \xi_0^{-2} - \gb k_F^2\langle f^2 \rangle/(4\pi\ef)$
\footnote{We neglect the small correction to $D(\q=0,\W\to 0)$ coming from the static part of $\Pi(\q=0,\W\to 0)$. Although this term scales as $(\kf\xi)^{-1}$ and  asymptotically wins over the $\xi^{-2}$ term, it has an  additional  smallness in $(\gb/\ef)^2$.}
Therefore
\begin{align}
 \label{eq:ind-scaling}
\delta \chi'_{ind} \propto \xi^{-2} \xi^2_0, \quad \delta \left(\frac{1}{\chi_{ind}}\right) \propto \xi^{-2},
\quad \Gamma_{ind} \propto \xi^2 \xi^4_0.
\end{align}

We emphasize that in, both cases, $\chi^\prime_{\xi =\infty} ( \W\to 0)$ remains finite
at the QCP because it differs from the thermodynamic
nematic susceptibility $\chi^\prime (\q\to 0, \W=0)$, which scales as $\xi^2$ and diverges at the QCP. We recall in this regard that we consider a clean system. In the presence of weak disorder the limits $\q \to 0$ and $\W \to 0$ indeed commute~\cite{Gallais2016}, but $\chi^\prime (0, \W)$ nonetheless approaches its clean limit behavior for frequencies above an appropriate transport scattering rate $\gamma_{tr}$.
In this respect, our $\chi^\prime (0, \W\to 0)$ is actually the susceptibility for $\W$ much larger than $\gamma_{tr}$, but well below any other energy scale.

{\it Comparison with experiments}~~~ The d-wave bosonic susceptibility is directly measured in polarization-resolved Raman scattering.
 The momentum transfer in Raman experiments is very low,  so that to high accuracy the susceptibility extracted from Raman measurements coincides with
$\chi (q=0, \Omega) = \chi (\W)$.
We compare our theoretical results for $\chi^{\prime\prime}(\W)$ with data for the Fe-chalcogenides FeSe/ FeSe$_{1-x}$S$_x$ (Refs. \onlinecite{Massat2016,Blumberg2017}) and Fe-pnictides BaFe$_2$As$_2$ (Ref. \onlinecite{Yang2014}) and NaFe$_{1-x}$Co$_x$As (Ref. \onlinecite{Thorsmolle2016}).
We assume that the data can be described within a clean limit, although disorder may be a source of systematic disagreement
~\footnote{The value $\gamma_{tr}$  at $T=0$ can be extracted  from the resistivity data. For FeSe we used the data from Ref. \cite{Huynh2014} and obtained  $\gamma_{tr} \sim 1 meV$. The data for  $Im \chi (0,\W)$ in Refs. \onlinecite{Blumberg2017,Thorsmolle2016} are for frequencies above roughly 3 meV, i.e. for all $\W$ the condition $\W > \gamma_{tr}$ is satisfied.}.
From the data one can extract the slope $\Gamma = \lim_{\Omega\rightarrow 0}\chi^{\prime \prime} (\W)/\W$.
Regardless of how the extrapolation to $\Omega\rightarrow 0$ is performed, $\Gamma$ grows rapidly in the vicinity of $T_S$.
~\footnote{
The data for $\Gamma$  are taken above a low frequency cutoff at ~3 meV, which obscures the divergence of $\Gamma$ very near $T_s$. We assume that $\Gamma(T=T_s)$ diverges and subtract a small temperature-independent constant from $\Gamma^{-1}$, as described in~\cite{suppl}.}.
The real part of the susceptibility $\chi^\prime(0,\W\to 0)$ was extracted~\cite{Massat2016,Blumberg2017,Thorsmolle2016}  from the data for $\chi^{\prime \prime} (\W)$  via Kramers-Kronig: $\chi^\prime(0,\W\to 0) = (2/\pi) \int_0^\infty d\W  \chi^{\prime \prime}(0,\W)/\W $}.

The nematic transition temperature $T_S$ varies with $x$ in FeSe$_{1-x}$S$_x$ and  NaFe$_{1-x}$Co$_x$As and vanishes at a particular S or Co doping.
We assume that the $T$ dependence can be incorporated into $\xi_0 (x,T)$ and $\xi (x,T)$,  but do not otherwise incorporate finite temperature into our calculations. As such the results should be valid as long as $\W > T_S$, which is true for most of the relevant experimental frequency range.
\begin{figure}
  \raggedright
  \includegraphics[width=\hsize,clip,trim=0 160 0 160]{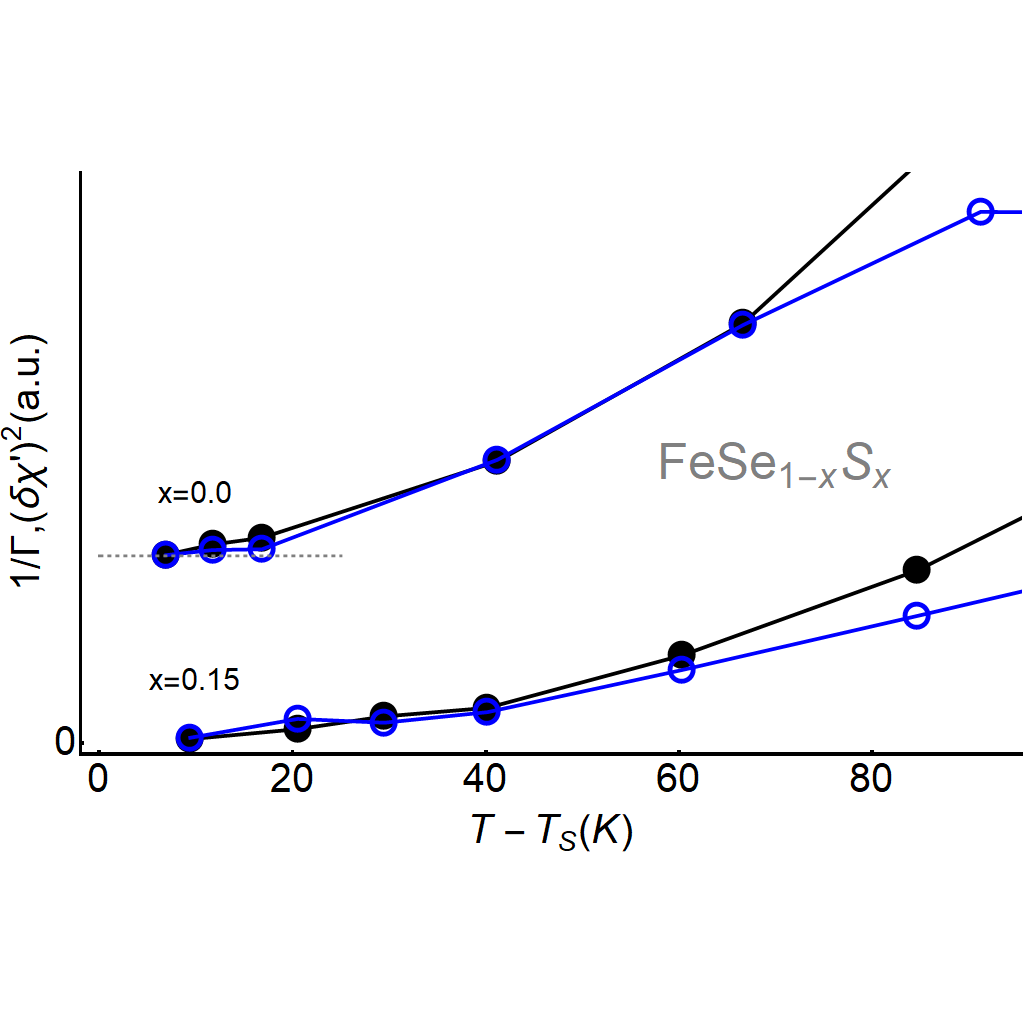}
  \hspace{-0.47\hsize}
  \llap{\raisebox{0.45\hsize}{
      \frame{
        \includegraphics[clip,trim=-15 170 -15 230,width=0.43\hsize]{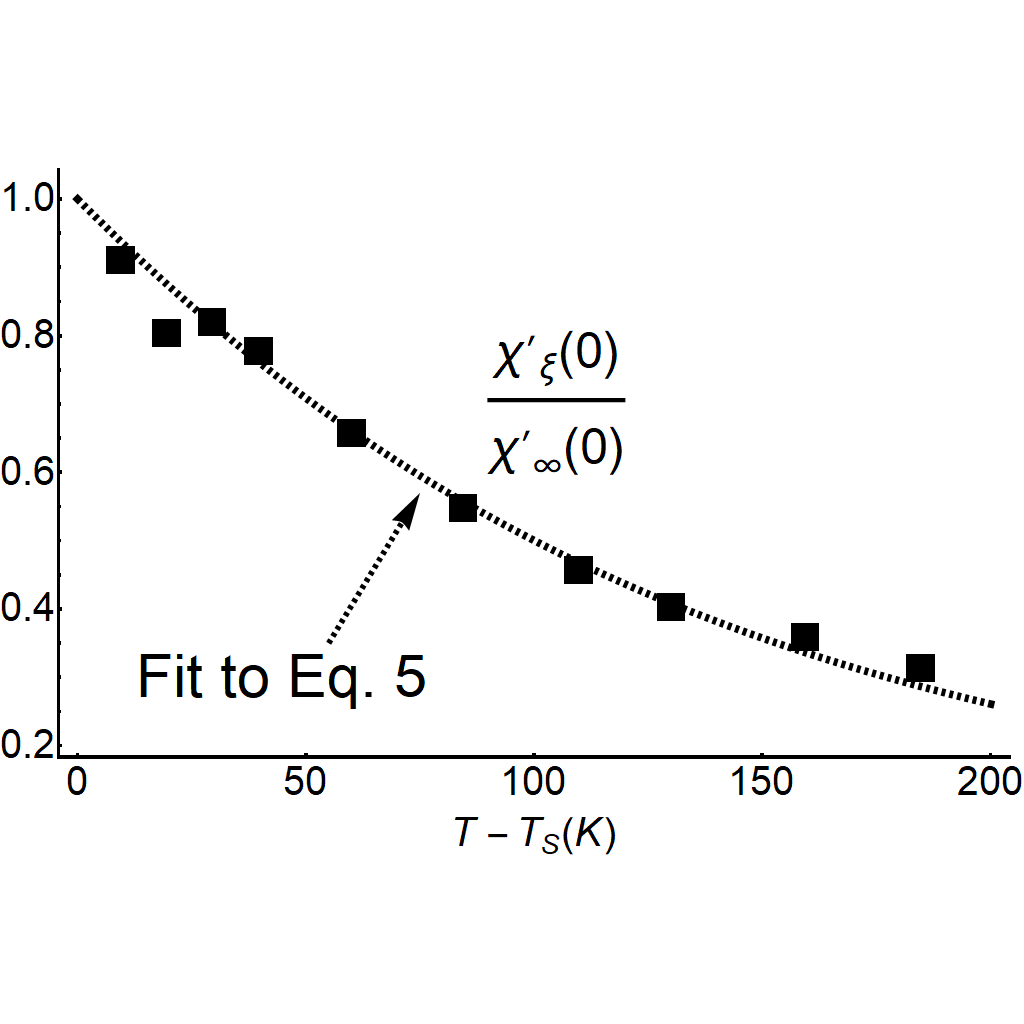}
      }}}
  \caption{
    $B_{2g}$ Raman data on FeSe$_{1-x}$S${_x}$ for dopings $x=0$ (shifted up for clarity) and $x=0.15$,  $(T-T_s)/T_s \lesssim 1$, taken from Ref. \onlinecite{Blumberg2017} [Similar data for $x=0$ are in Ref.~\onlinecite{Massat2016}].
    Filled circles -- $1/\Gamma$, where $\Gamma = \chi^{\prime \prime}(0, \W)/\W$ is the slope of measured
   $\chi^{\prime \prime} (0,\W)$ at small frequencies. Open circles -  $(\delta \chi^\prime)^2 = (\chi_{\xi = \infty}^\prime (\Omega \to 0) - \chi_{\xi}^\prime (\Omega \to 0))^2$, where $\chi_{\xi}^\prime (\Omega \to 0)$ has been obtained from $\chi^{\prime \prime}(0, \W)$ by KK transform.
    The data for $(\delta\chi')^2$ have been rescaled by a constant factor \cite{suppl}. The data show that $(\delta \chi^\prime)^2$ and $1/\Gamma$ scale together, i.e., their ratio is independent of $\xi$. Such behavior is consistent with the Pomeranchuk scenario described in the text.
    The inset shows the experimental $\chi_{\xi}^\prime (\Omega \to 0)$ along with the fit to Eq. (\ref{eq:finite-A}) using $\xi (T)$ extracted from the data for $\Gamma$ \cite{suppl}.}
  \label{fig:FeSe-fits}
\end{figure}

For both sets of materials we examined possible scaling between $1/\Gamma$ and powers of $\delta\chi^\prime$. For FeSe$_{1-x}$S$_x$ we found that $(\delta \chi^\prime)^2$ and $1/\Gamma$ scale together (Fig. \ref{fig:FeSe-fits}). Such behavior is consistent with the Pomeranchuk scenario, as in this case both $(\delta \chi^\prime)^2$ and $1/\Gamma$ scale as $\xi^{-2}$, Eq. (\ref{eq:pom-scaling}).
The data for FeSe$_{1-x}$S$_x$ also show~\cite{Blumberg2017} that $\chi^\prime_{KK} (0,0)$ increases as the system approaches the nematic transition but
 deviates from Curie-Weiss behavior
near the transition point.
The deviation gets more pronounced with increasing $x$. Such behavior is also consistent with the Pomeranchuk scenario, Eq. (\ref{eq:finite-A}), particularly given that $k_F$ in this system is small for all pockets~\cite{Coldea2017,Terashima2014,Charnukha2016}), because in this case $\chi^\prime (0,\W \to 0)$ increases as $\xi^2$ between $\xi = O(1)$ and $\xi \sim 1/k_F$.
The data also show~\cite{Massat2016,Blumberg2017} that the maximum in $\chi^{\prime \prime} (0,\W)$ remains at a nonzero frequency at the nematic transition.
This is consistent with the crossover to QC behavior because at $\xi = \infty$, $\chi^{\prime \prime} (0,\W)$ still increases at small $\W$ as $\W^{1/3}$, and therefore passes through a maximum at nonzero $\W$. We consider the combination of these results  as a strong indication that nematicity in FeSe/FeSe$_{1-x}$S$_x$ is caused by a d-wave Pomeranchuk instability.

\begin{figure}
  \raggedright
  \includegraphics[width=\hsize,clip,trim=0 160 0 160]{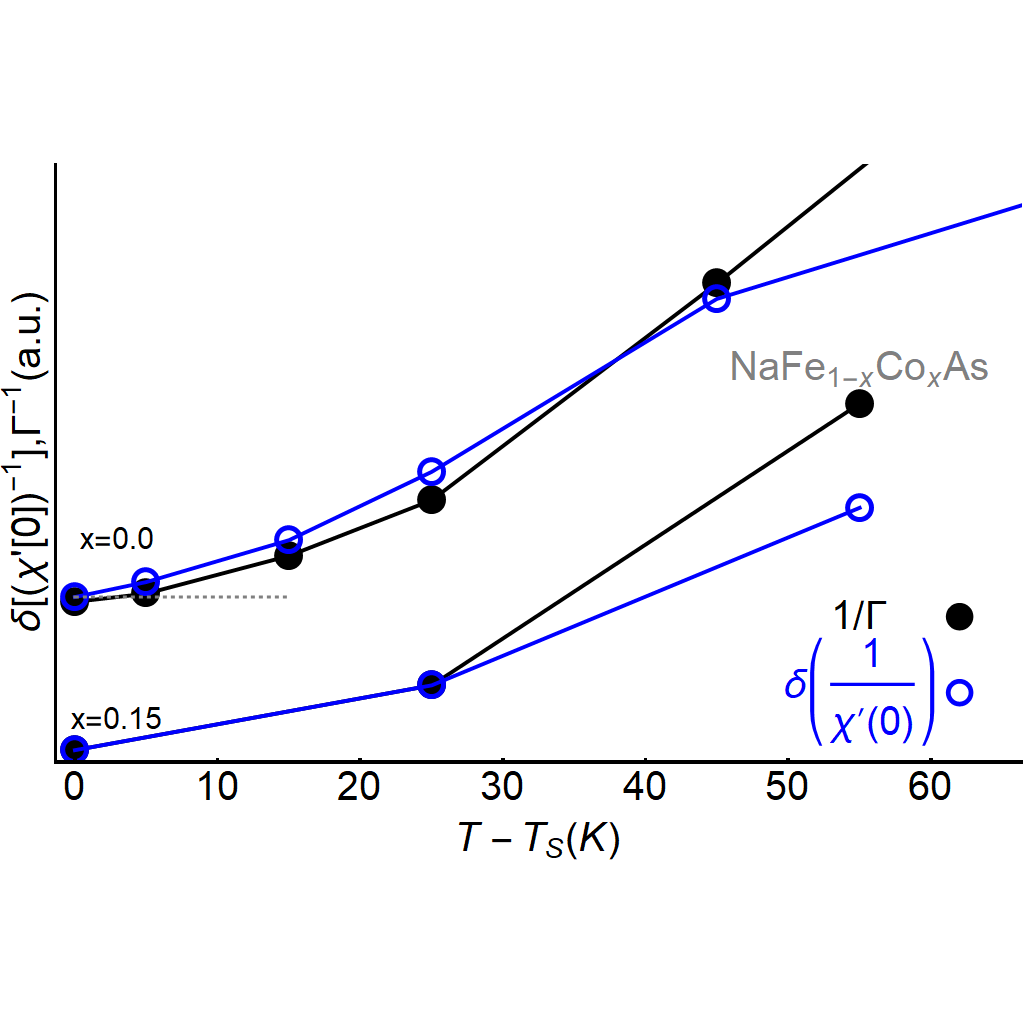}
  \hspace{-0.545\hsize}
  \llap{\raisebox{0.43\hsize}{
      \frame{
        \includegraphics[clip,trim=-15 170 10 230,width=0.4\hsize]{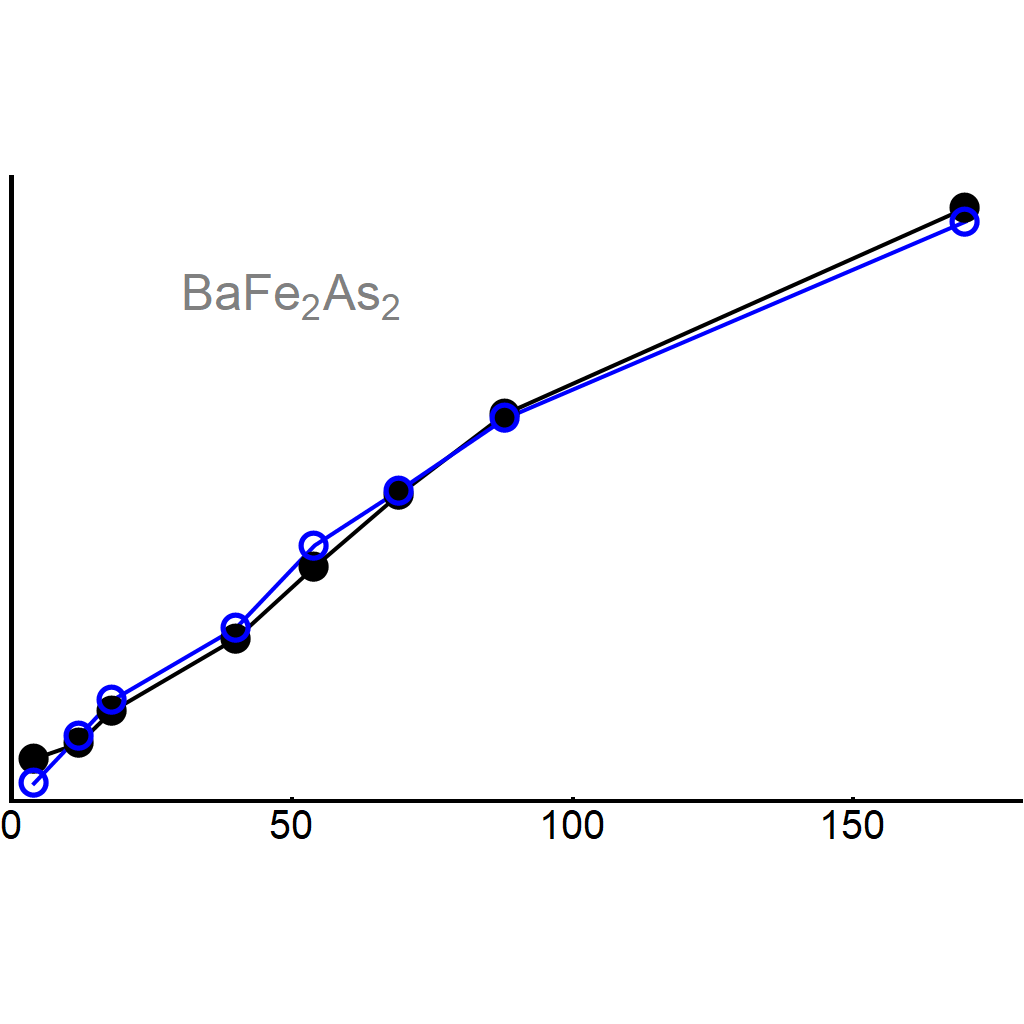}
      }}}
  \caption{
    $B_{2g}$ Raman data for NaFe$_{1-x}$Co$_{x}$As for dopings $x=0$  (shifted up for clarity) and $x=0.15$, $(T-T_s)/T_s \lesssim 1$ (taken from Ref. ~\onlinecite{Thorsmolle2016}).
    Filled circles -- $1/\Gamma$, open circles -- $\delta (1/\chi^\prime)$.
    Inset-- the same quantities for undoped Ba122 (data from Ref. \onlinecite{Yang2014}). The data show that $\delta (1/\chi^\prime)$ and $1/\Gamma$ scale together in both NaFe$_{1-x}$Co$_{x}$ and BaFe$_2$As$_2$.  Such behavior is  consistent with the independent scenario at
    $|T - T_s| \ll T_s$, but the fact that the scaling holds in a wide range of $T$  requires further study.}
  \label{fig:NaFeCo-fits}
\end{figure}

For Ba$($FeAs$)_2$ and NaFe$_{1-x}$Co$_x$As, we found that the temperature dependence of $1/\Gamma$  closely follows that of $\delta(1/\chi)$ over several tens of kelvin near $T_S$, as shown in Fig. \ref{fig:NaFeCo-fits}. This observation is generally consistent with the ``independent'' scenario as there both $1/\Gamma$ and $\delta(1/\chi)$ scale as $\xi^{-2}$ (see Eqs. (\ref{eq:ind-scaling})).   A natural candidate for the independent order parameter is the composite Ising nematic operator derived from the magnetic order parameter~\cite{Fernandes2012}, since in these materials the magnetic and nematic transitions are close to each other and show nearly identical doping dependence. We caution, however, that the scaling $1/\Gamma\sim  \delta(1/\chi)$ holds in our theory under the assumption that  $\xi_0$ is essentially a constant, and hence $\chi^\prime \propto \xi^2_0$ is also a constant.  This is the case near $T_s$, but at higher $T$, the measured $\chi^\prime$ varies significantly over the temperature ranges shown in Fig. \ref{fig:NaFeCo-fits} (Refs. \cite{Yang2014,Thorsmolle2016}). More data in the range $T -T_S < T_S$, and a more careful analysis (possibly accounting for the effect of impurities) is needed to fully understand the data for Ba122 and Na111.

{\it Summary}~~~
In this work we analyzed the dynamic response of a clean 2D system of itinerant fermions coupled to a scalar boson $\phi$, which undergoes a continuous transition towards a d-wave charge nematic order. We obtained the form of $\Pi (0, \W)$  both at and near the nematic transition and related it to the bosonic susceptibility $\chi (0, \Omega)$ in the case where $\phi$ is a soft collective mode of fermions near a Pomeranchuk instability and when it is an independent critical degree of freedom, such as a composite spin order parameter near an Ising-nematic transition. We compared our results with  polarization-resolved Raman data  for FeSe$_{1-x}$S$_x$ and BaFe$_2$As$_2$/NaFe$_{1-x}$Co$_x$As. We argued that the data for FeSe$_{1-x}$S$_x$, which does not order magnetically down to $T =0$, are
well described by the d-wave charge Pomeranchuk scenario. The data for BaFe$_2$As$_2$/NaFe$_{1-x}$Co$_x$As at $T \geq T_s$ are consistent with the independent boson scenario (for which composite spin order is the primary candidate), but further study is required to understand the data at higher $T$.
For all compounds we found evidence for quantum critical behavior near $T_S$. We hope that our findings  will motivate further measurements
in the range $T - T_S < T_S$ to allow for quantitative comparison of theoretical predictions with experiment.

Our analysis neglects interaction with acoustic phonons. This interaction does not directly affect $\chi (q=0,\W)$, but it does contribute to $\chi (q \to 0, \W=0)$~\cite{Onari2012,Gallais2016}) and therefore gives an additional contribution to the difference between $\xi_0$ and $\xi$. It also affects the momentum dependence of $D(q,\W)$ at $\W < v_F q$, and eventually cuts off the critical behavior at $T=T_s$~\cite{Karahasanovic2016,Paul2017}.  Given that $\Gamma$ and $\chi^\prime$ strongly increase as $T$ approaches $T_s$, we conjecture that coupling to acoustic phonons affects the system's
 dynamics only in a narrow range very near $T_S$, while our theory is applicable outside this range.

{\it Acknowledgements}~~~  We thank G. Blumberg, R. Fernandes, Y. Gallais,  M. Navarro-Gastiasoro, I. Paul, J. Schmalian, M. Tanatar, and V. Thorsmolle
 for stimulating discussions. We are thankful to G. Blumberg,
W.-L. Zhang and V. Thorsmolle
for sharing unpublished data with us. This work was supported  by the NSF DMR-1523036 (AK and AVC). SL and DC are supported by a postdoctoral fellowship from the Gordon and Betty Moore Foundation, under the EPiQS initiative, Grant GBMF-4303, at MIT.
 AVC and DC acknowledge the hospitality of the Aspen Center for Physics,  which is supported by NSF grant PHY-1607611. AK, DC, EB, and AVC acknowledge the hospitality of KITP at UCSB,  which is supported by NSF grant PHY-1125915.
 \bibliography{QCP,QCP_AC,NZS}
 
\clearpage
\onecolumngrid
\appendix

\section{Supplementary material}

In this Supplementary material we detail the methods we used to fit our analytic expressions for the dynamical polarization for a non-conserving order parameter to the experimental data that appear in Figs. \ref{fig:FeSe-fits} and \ref{fig:NaFeCo-fits} and discuss thew role of disorder.

\subsection{Details of the fitting procedure}
\label{sec:suppl-mater-deta}

The expression for the dynamical polarization was given in Eqs.~(68)-(70) of Ref. \onlinecite{Klein2017a}. An approximate version of Eq.~(70) in that paper can be written as:
\begin{equation}
  \label{eq:dyn-int-qcp-zero-temp}
  \gb\Pi(\q = 0,\W_m) \approx \kf^2\left(\frac{\gb}{\ef}\right)^2\langle f_2\rangle\int_0^{\W_m}\frac{d\W}{\W_m}\int_0^{1}\frac{d\theta}{\tp}\frac{|\theta|^2}{|\theta|^2 + \frac{\gamma\langle f^2\rangle |\W_m+\W|}{\theta\kf^3\vf} + \left(\frac{(\W+\W_m)}{\bar \Omega}\right)^2 +(\kf\xi)^{-2}}.
\end{equation}
Here $\theta$ is the  momentum component along the Fermi surface rescaled by $k_F$, $f(\theta)$ is a  nematic form factor,  $\xi$ is the correlation length, other parameters are some constants. Setting $\W_m =0$ and performing frequency integration and the integration over $\theta$,  we obtain Eq. (\ref{eq:finite-A}). Keeping $\W_m$ finite, subtracting from Eq. (\ref{eq:dyn-int-qcp-zero-temp}) the static part, and again performing integrations over $\Omega$ and over $\theta$, we obtain Eq. (\ref{eq:finite-C}). Note that the upper limit of integration over  is in general of order one. We set it equal to one for simplicity. However, the choice of the upper limit of the $\theta$ integration does not affect the term linear in $\xi^{-1}$ in Eq. (\ref{eq:finite-A}) and the frequency dependence in Eq. (\ref{eq:finite-C}) in the main text.

\begin{figure}
  \centering
  \subfloat[]{\includegraphics[width=0.48\hsize,clip,trim=50 160 0 160]{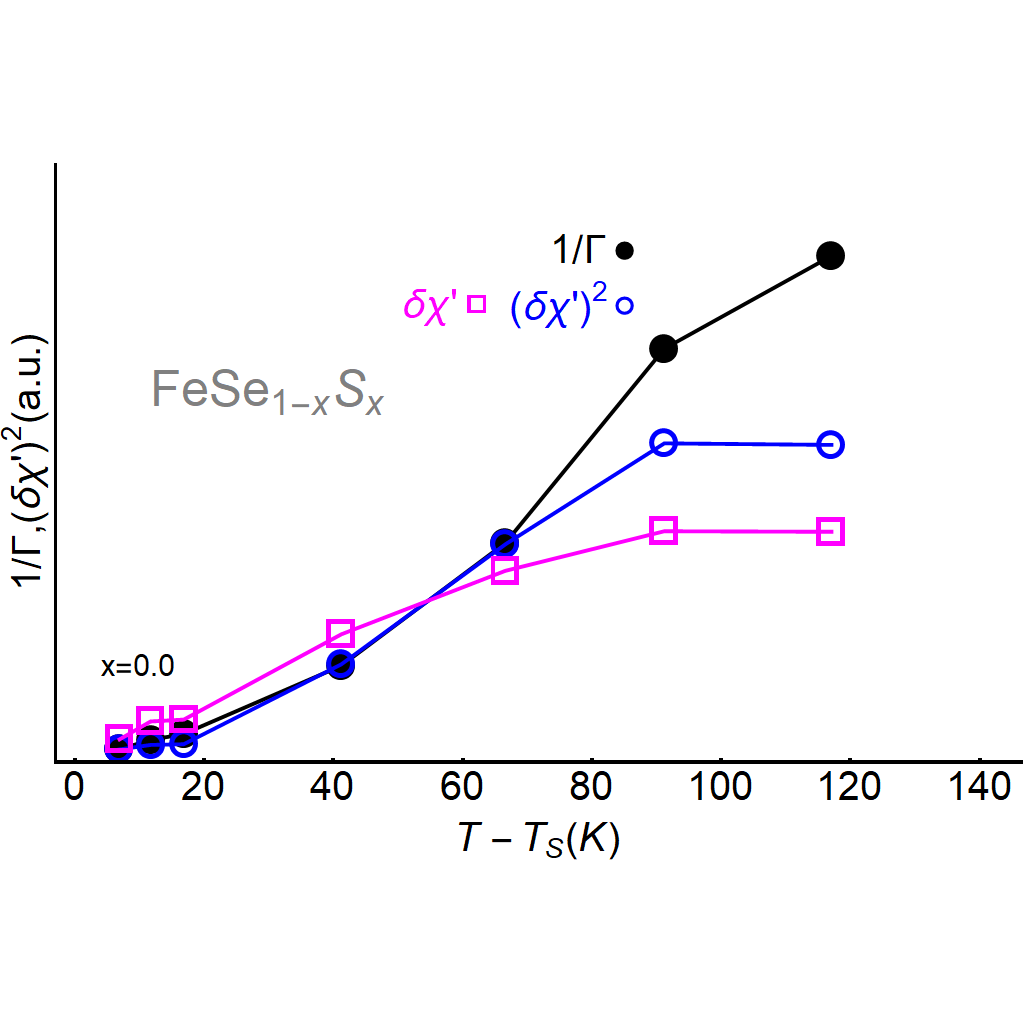}}
 \quad \subfloat[]{\includegraphics[width=0.48\hsize,clip,trim=50 160 0 160]{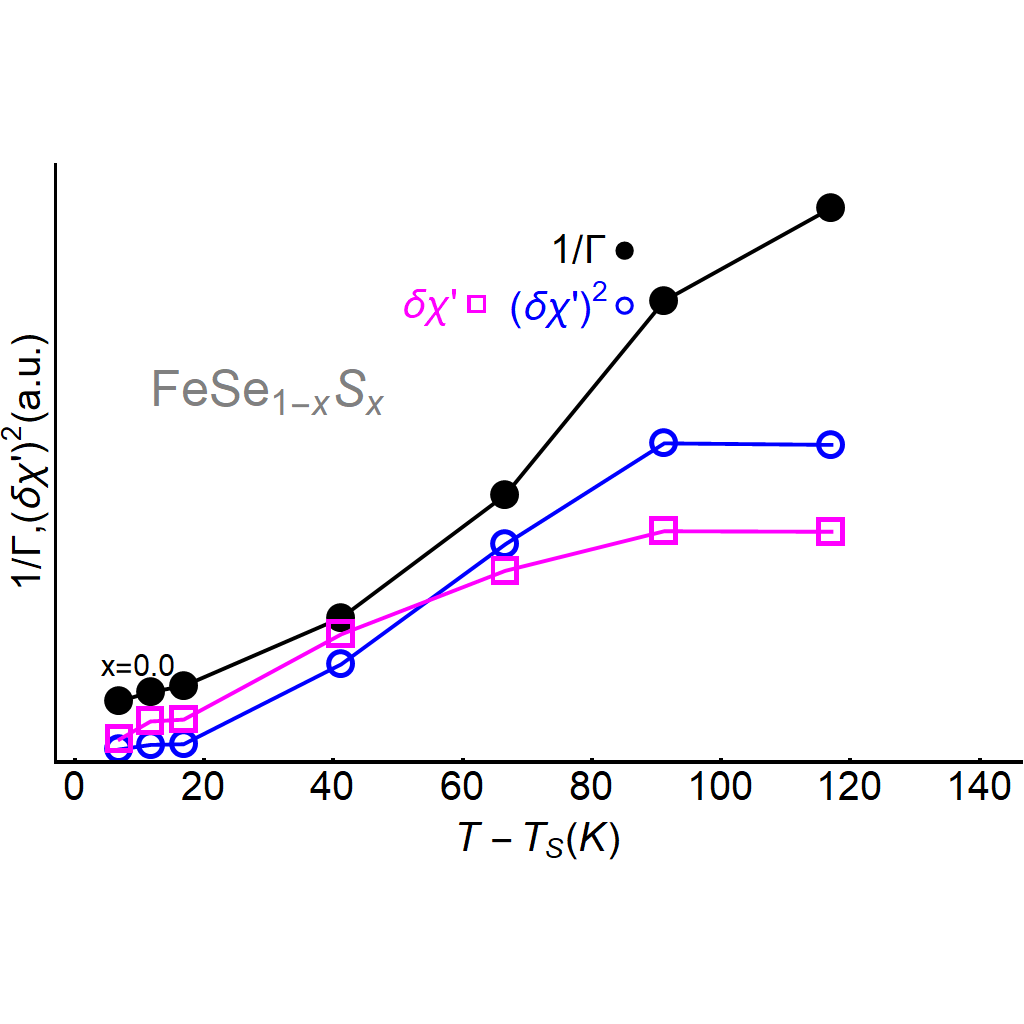}}\\
 \subfloat[]{\includegraphics[width=0.48\hsize,clip,trim=50 160 0 160]{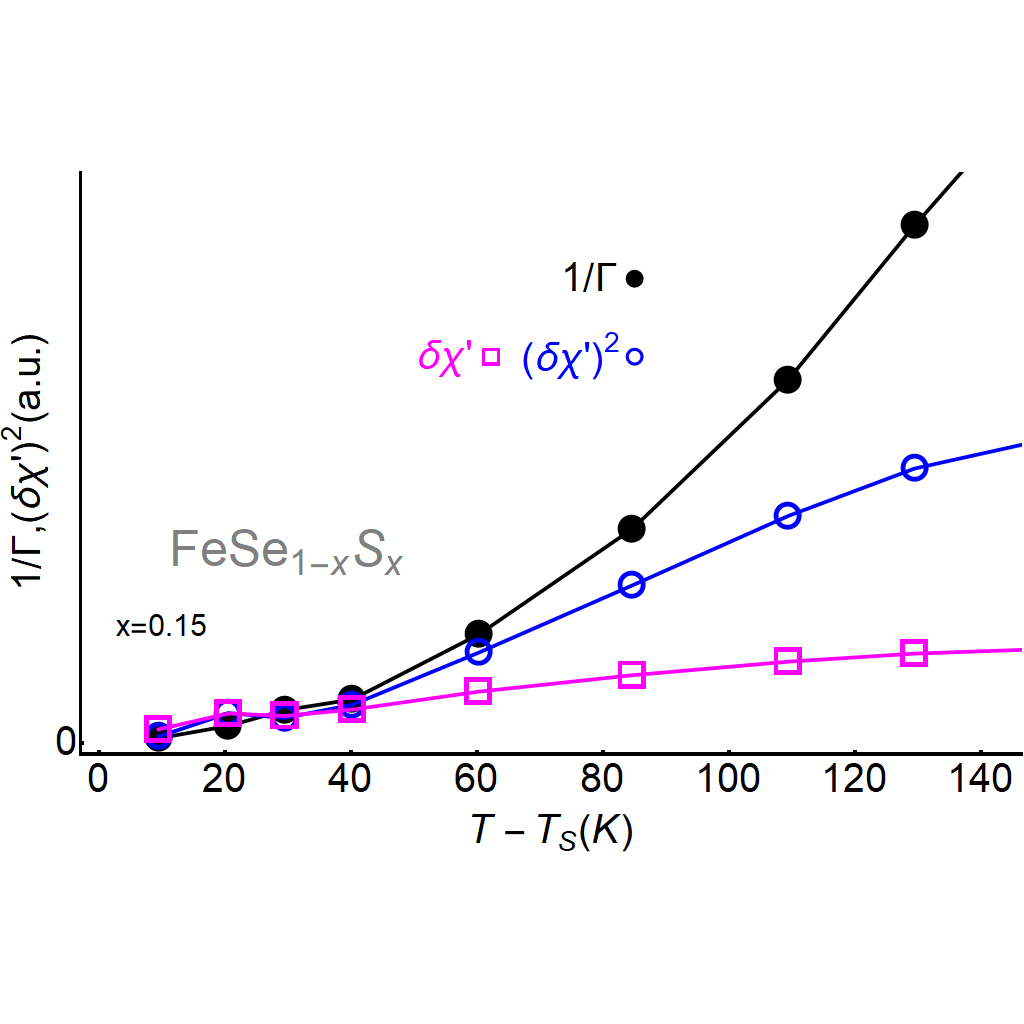}}
 \quad \subfloat[]{\includegraphics[width=0.48\hsize,clip,trim=50 160 0 160]{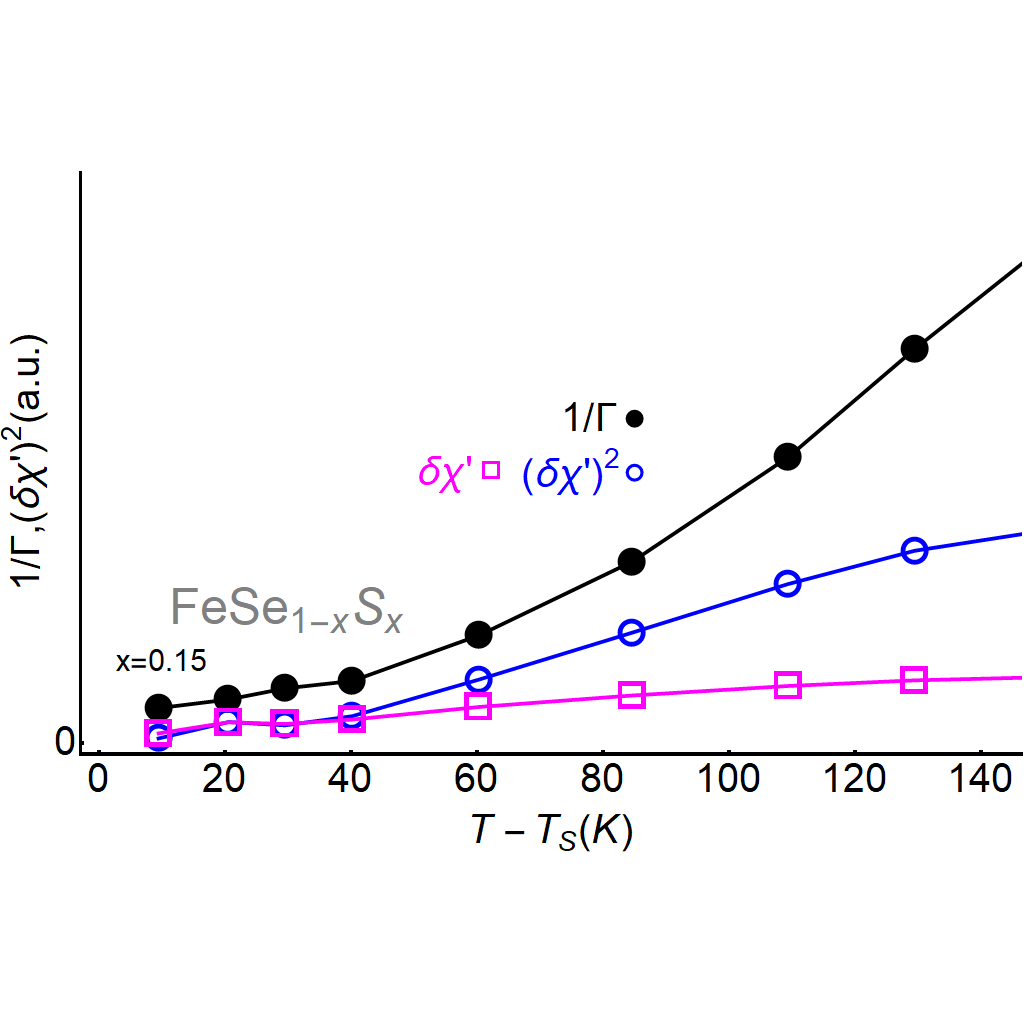}}
   \caption{(a) The scaling of $\Gamma^{-1}$ with $\delta\chi',(\delta\chi')^2$ for FeSe$_{1-x}$S$_x$, at $x=0$. (b) The same scaling plot, but with the small constant $\Gamma_0^{-1}$ restored. (c),(d) -- Same as in Figs. (a) and (b), but for $x=0.15$.}
  \label{fig:supp-fese}
\end{figure}

We obtained  the data points for Fig. \ref{fig:FeSe-fits} for FeSe/FeSe$_{1-x}$S$x$ from the authors of Ref. \onlinecite{Blumberg2017}. In producing the figure, we removed a small background term in the data for $\Gamma^{-1}$, i.e. set
\begin{equation}
  \label{eq:gamma-fit}
  \Gamma^{-1} = \Gamma_{data}^{-1} - \Gamma^{-1}_0.
\end{equation}
We attribute this small background term to the finite frequency cutoff in the Raman data at about 3 meV, and the extrapolation procedure to zero frequency. However, the scaling of $\Gamma^{-1}$ with $\delta\chi^2$ holds even if we don't remove $\Gamma^{-1}_0$. The  existing data for $\chi'(0)$ are for $T \geq T_s$, with no data point right at $T=T_s$.
We  extrapolated the data to $T_S$ and obtained $\delta\chi = \chi'_{T=T_S} - \chi'_T $. Again, the  result for $\delta \chi'$ does not change significantly if we subtract from $\chi'_T$ its value at $T$ closest to $T_s$ instead of using extrapolation. Finally, in comparing the scaling curves in Figs. 2 and 3 in the main text, we rescaled the data for one quantity to match the other one at the lowest temperatures. To obtain the inset in Fig. 2, we extracted $\xi$ from the data on $\Gamma$ by just using  $\Gamma^{-1} \propto \xi^2$, without assuming any particular $T$ dependence of $\xi$. We fitted $\Gamma^{-1}$ to a power law and obtained an excellent agreement with the data  for $\xi^{-1} = a (T- T_S)$, where $a$ is a constant.  We used Eq. (\ref{eq:finite-A}) with $a$ as a single fitting parameter for the  plot in the inset of Fig. 2.
For a more complete view, we refer to Fig. \ref{fig:supp-fese}, where we plot $1/\Gamma$ and either  $\delta\chi'$ or $(\delta\chi')^2$.  This figure shows that for FeSe/FeSe$_{1-x}$S$x$, $1/\Gamma$ scales with $(\delta\chi')^2$, but not with  $\delta\chi'$.

\begin{figure}
  \centering
  \subfloat[]{\includegraphics[width=0.48\hsize,clip,trim=50 160 0 160]{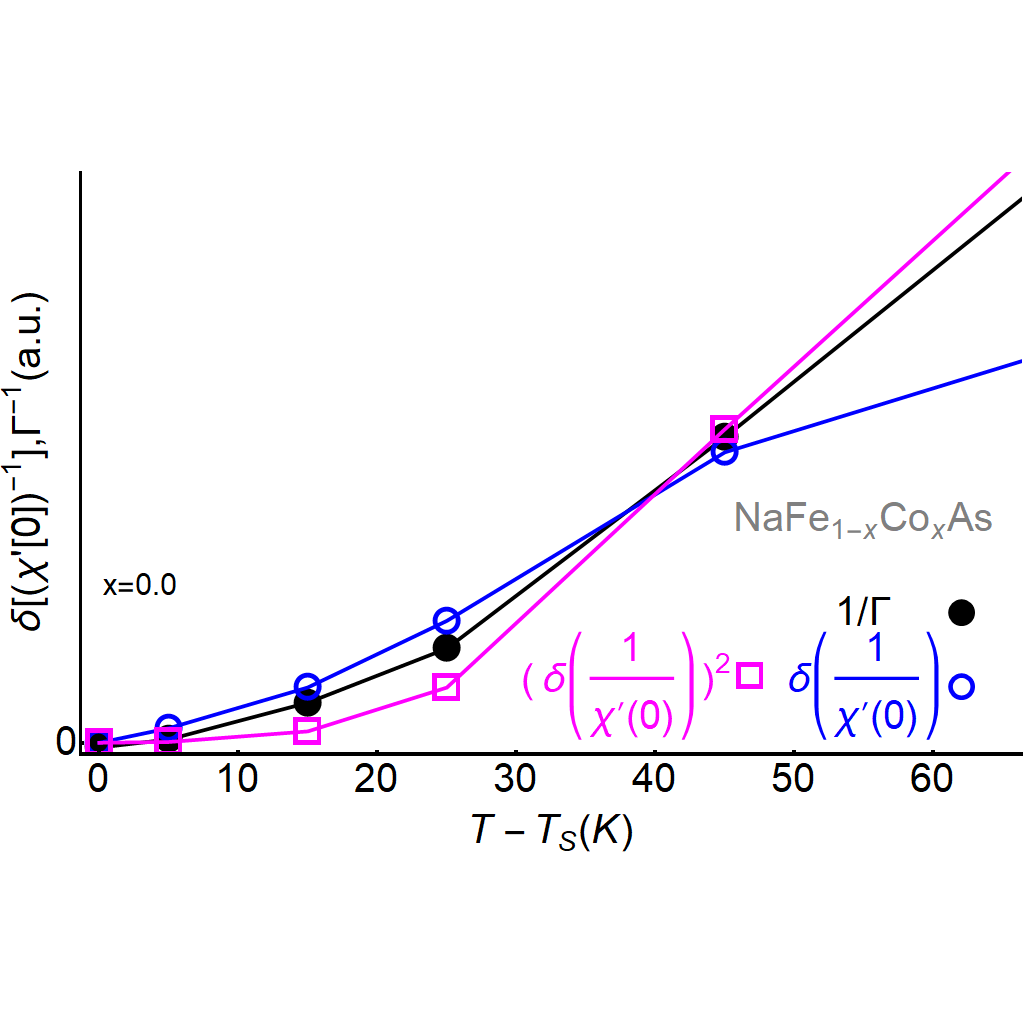}}
 \quad \subfloat[]{\includegraphics[width=0.48\hsize,clip,trim=50 160 0 160]{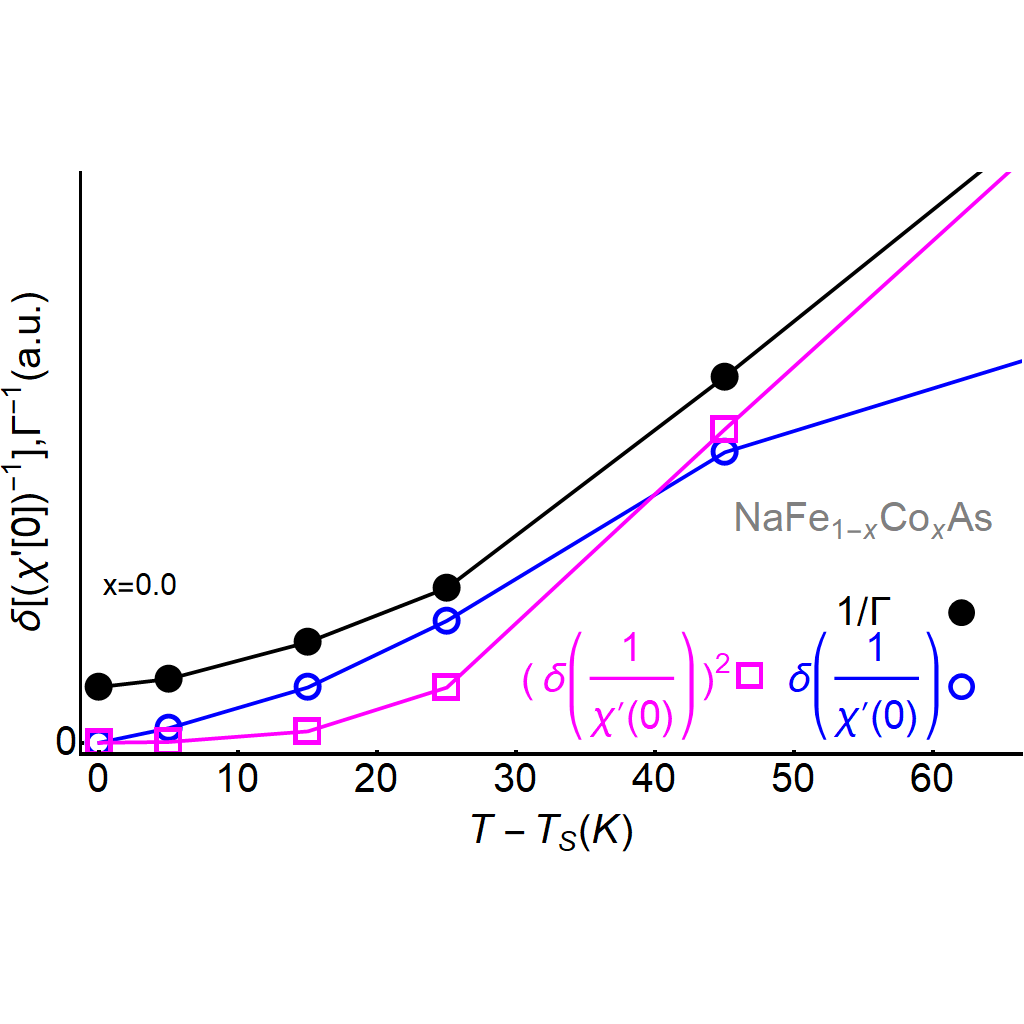}}\\
 \subfloat[]{\includegraphics[width=0.48\hsize,clip,trim=50 160 0 160]{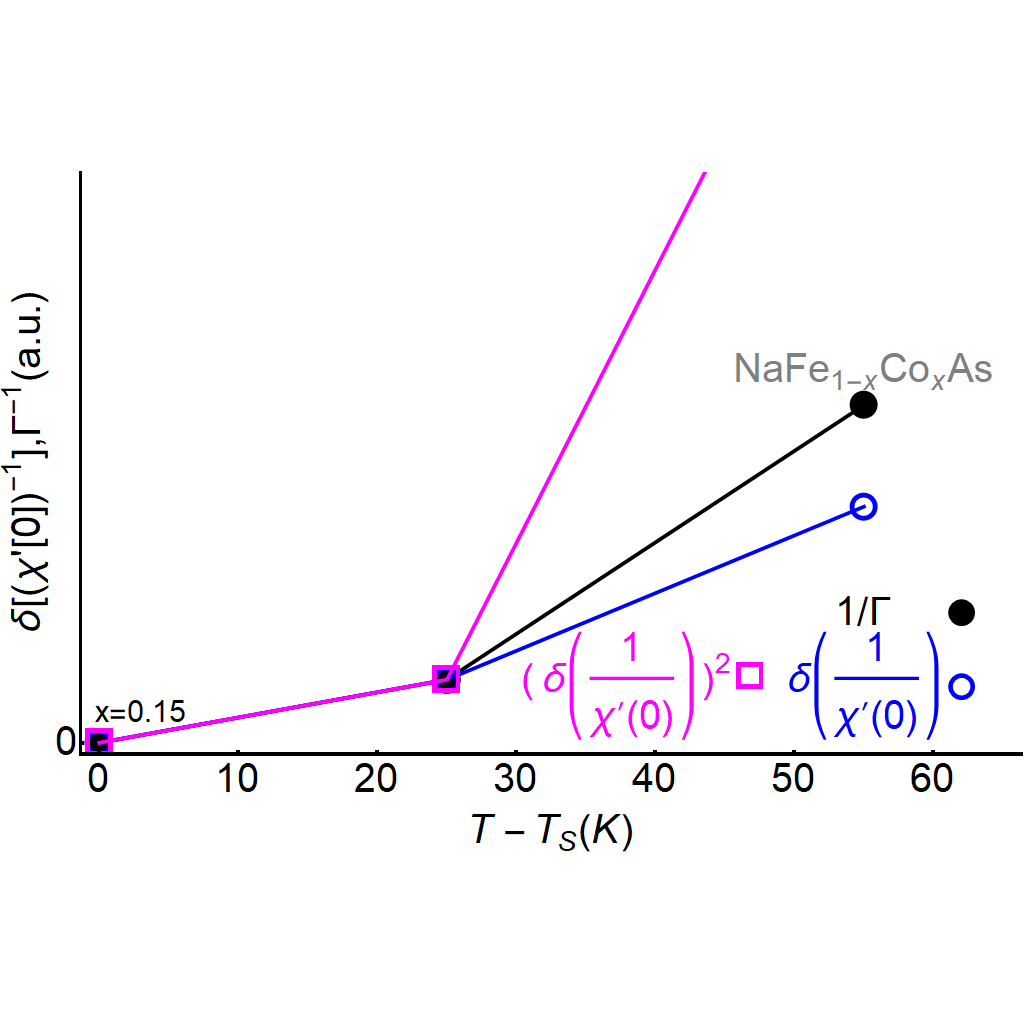}}
 \quad \subfloat[]{\includegraphics[width=0.48\hsize,clip,trim=50 160 0 160]{chi_0_vs_peak_Na111_supp21}}
  \caption{The scaling of $\Gamma^{-1}$ with $\delta(1/\chi'), \left(\delta(1/\chi')\right)$ for NaFe$_{1-x}$As$_x$. The figures are arranged in the same manner as in Fig. \ref{fig:supp-fese}.}
  \label{fig:na111-supp}
\end{figure}

\begin{figure}
  \centering
\subfloat[]{\includegraphics[width=0.48\hsize,clip,trim=42 160 0 160]{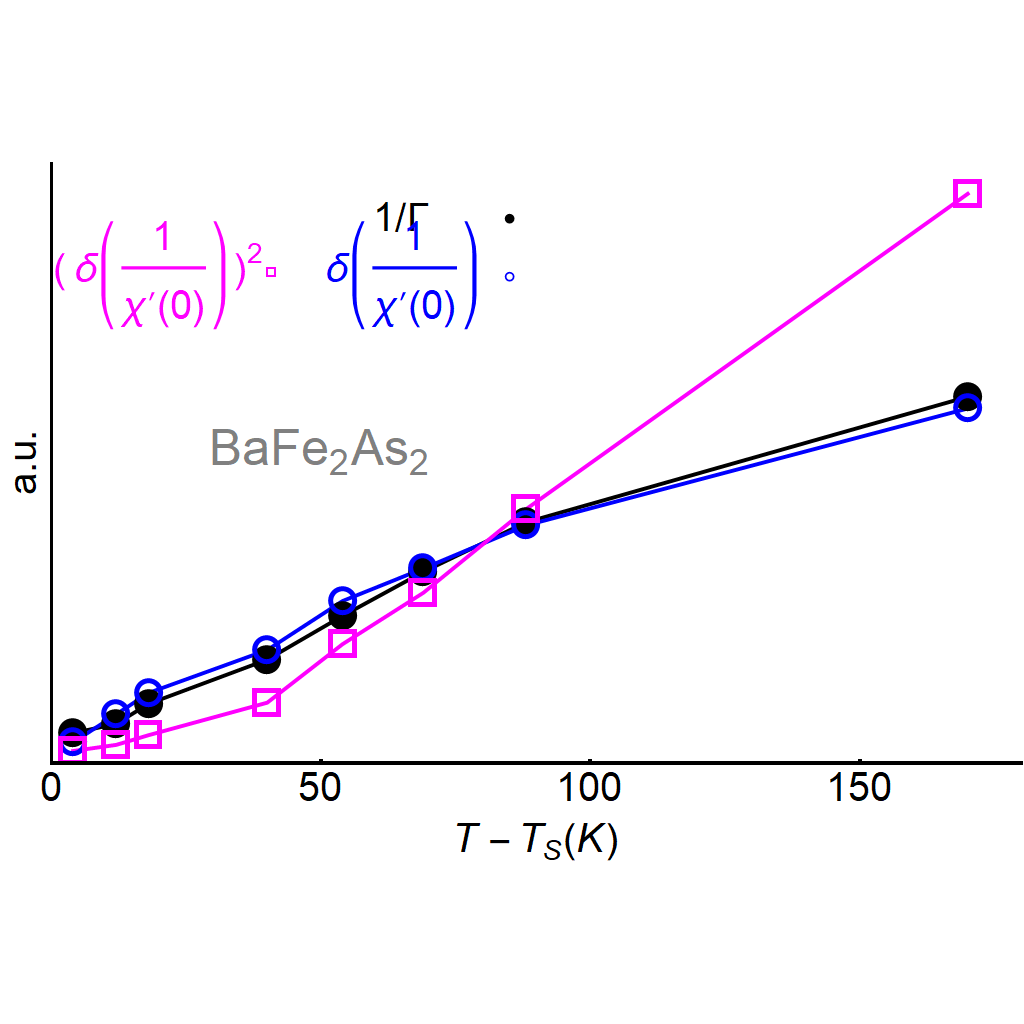}}
 \quad \subfloat[]{\includegraphics[width=0.48\hsize,clip,trim=42 160 0 160]{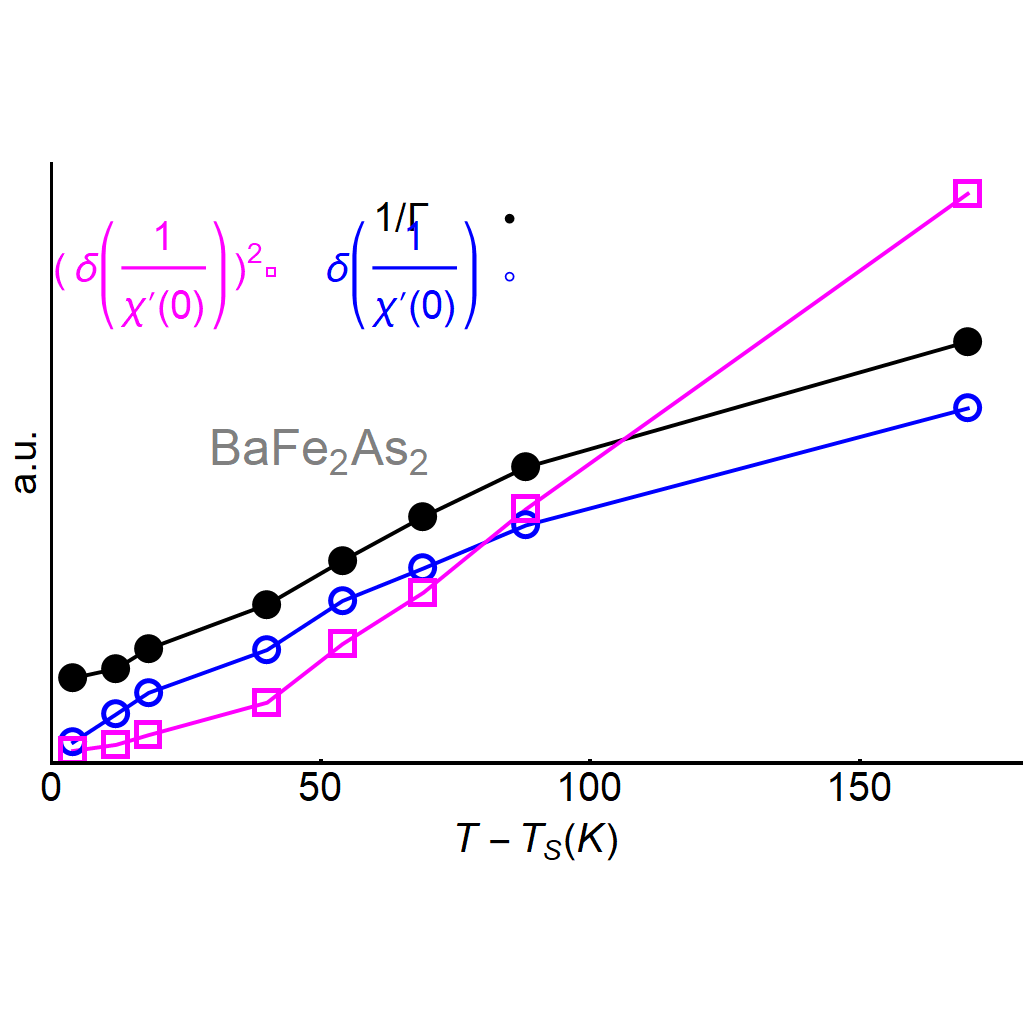}}
 \caption{The scaling of $\Gamma^{-1}$ with $\delta(1/\chi'), \left(\delta(1/\chi')\right)$ for BaFe$_2$As$_2$. The figures are arranged in the same manner as in Fig. \ref{fig:supp-fese}.}
 \label{fig:ba122-supp}
\end{figure}

The data points in Fig. \ref{fig:NaFeCo-fits} were taken from Figs. 3 and 7d in Ref. \onlinecite{Thorsmolle2016}. The data points in the inset were taken from Fig. 1c of Ref. \onlinecite{Yang2014}. The analysis of the data was done in the same way as for Fig. 2.  In Fig. \ref{fig:na111-supp} we plot $\Gamma^{-1}$ and either  $\delta(1/\chi')$ or $\left(\delta(1/\chi')\right)^2$ for the Na111 compound. In Fig. \ref{fig:ba122-supp} we show the same for Ba122. We clearly see that for  Na111 and Ba122,  $\Gamma^{-1}$ scales with $\delta(1/\chi')$  but not $\left(\delta(1/\chi')\right)^2$.

All data points (for both figures) were extracted using Engauge Digitizer \cite{engauge}.

\subsection{The role of disorder}
\label{sec:suppl-mater-brief}

In this subsection we briefly discuss the effects of weak disorder. In the presence of disorder, $\Pi (q=0, \W)$ becomes non-zero even in the absence of interaction, and at the smallest momentum and frequency, $\Pi (q, \W)$ becomes a regular function of $q$ and $\W$. At zero momentum, $\Pi (q=0, \W) \propto  \W \tau_{tr}/(1 + (\W \tau_{tr})^2)$, where $\tau_{tr} =1/\gamma_{tr}$ is a transport lifetime due to impurities.  Substituting this $\Pi$ into the expression for the bosonic susceptibility for the
 ``independent boson'' scenario, one obtains~\cite{Gallais2016}
\begin{equation}
  \label{eq:supp-imp-1}
  \chi_{ind} = \frac{1}{\xi_0^{-2}-i \alpha \W/\gamma_{tr}}
\end{equation}
where $\alpha$ is an appropriate coupling constant. In this situation both $\Gamma,\chi'(0)$ increase but do not diverge at $T_s > T_0$, where again $T_0$ is the transition temperature of the $\phi$ field. ($T_s$ can still be larger than $T_0$ due to coupling to additional degrees of freedom such as phonons.) In such a model, $\Gamma \propto \xi_0^4/\gamma_{tr}$ scales with $\chi' \propto \xi_0^2$ rather than  with $\delta\chi'$:  $\Gamma \sim \left( \chi'(0) \right)^2$.

\begin{figure}
  \centering
\subfloat[]{\includegraphics[width=0.48\hsize,clip,trim=0 60 0 60]{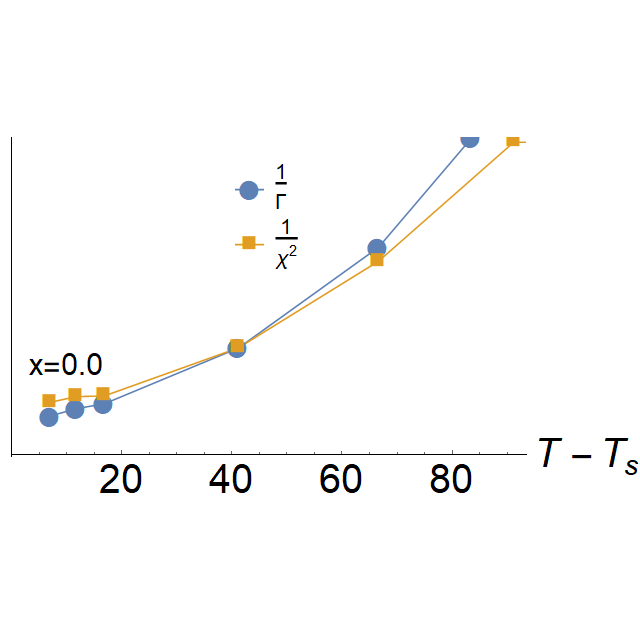}}
 \quad \subfloat[]{\includegraphics[width=0.48\hsize,clip,trim=0 60 0 60]{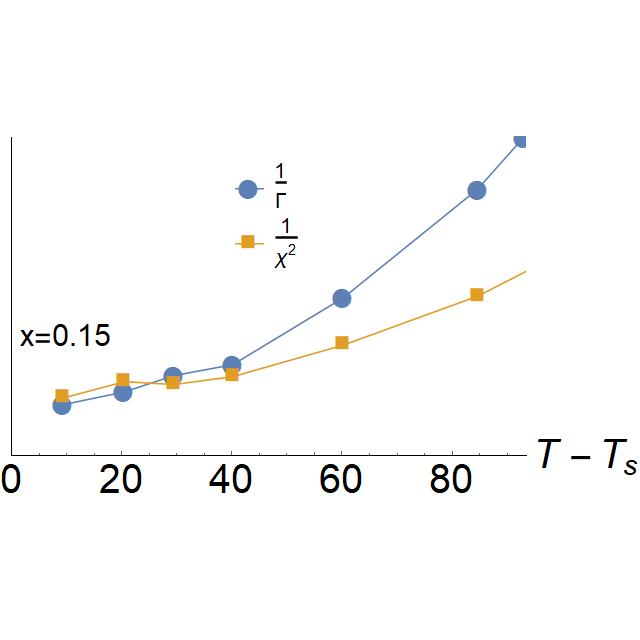}}
 \caption{The scaling of $\Gamma^{-1}$ with $(\chi')^2$ for FeSe, without background removal, consistent with a theory of an independent order parameter with disorder. The data do not scale well, and the disagreement is stronger at higher doping.}
 \label{fig:FeSeimp-supp}
\end{figure}

In the presence of interaction,  the contribution to the slope of $\Pi (q=0,\W)$ from the interaction scales as $\xi^2$ and is larger than the one due to disorder.  However, at a finite $\xi$, the interplay between the contributions to $\Pi$ from the disorder and from interaction depends on system parameters and may vary from one system to the other.
We checked the possibility that the scaling in  FeSe$_{1-x}$S$_x$ may be due to disorder by comparing the behavior of $\Gamma$ with that of $\left( \chi'(0) \right)^2$.  We plot the two quantities in Fig. \ref{fig:FeSeimp-supp}. We believe the data do not agree well with the disorder scenario for three reasons: (a) For x=0.15, there is strong disagreement between the two curves; (b) For x=0, the data for $1/\Gamma$ and $1/\chi'(0)^2$ cross at a finite $T - T_s$. This implies that the two curves do not cross zero at a single transition temperature $T_0$, as would be expected if both were proportional to $\xi^{-4}_0$; (c) For the doped compound, the disagreement between the two curves is stronger than for the undoped compound. Such behavior is counter to the naive expectation that disorder should grow stronger with doping.
For the Na111 and Ba122 materials, the agreement between the data and the disorder scenario is better, and, as we said in the main text, more data are needed to fully understand the origin of the scaling behavior in Na111 and Ba122 materials.

\end{document}